\documentclass[aps,showpacs]{revtex4}
\usepackage{amssymb}
\usepackage{graphicx}
\usepackage[english]{babel}
\usepackage{indentfirst}
\usepackage{amsxtra}
\usepackage{amsmath}
\usepackage{supertabular}
\usepackage{multirow}
\usepackage [mathcal]{eucal}
\newcommand{\beq}{\begin{equation}}
\newcommand{\eeq}{\end{equation}}
\newcommand{\beqn}{\begin{eqnarray}}
\newcommand{\eeqn}{\end{eqnarray}}
\newcommand{\sumint}{\sum \!\!\!\!\!\!\!\!\int }
\newcommand{\bsigma}{\mbox{\boldmath $\sigma$}}
\newcommand{\btau}{\mbox{\boldmath $\tau$}}

\newcommand{\br}{{\bf r}}

\newcommand{\bs}{{\bf s}}
\newcommand{\bl}{{\bf l}}
\newcommand{\bN}{{\bf \nabla}}
\newcommand{\bY}{{\bf Y}}

\newcommand{\oxy}{$^{16}$O~}

\newcommand{\bmon}{$B({\cal M}1)\!\!\uparrow$~}
\newcommand{\bmtw}{$B({\cal M}2)\!\uparrow$~}
\newcommand{\bmth}{$B({\cal M}3)\!\uparrow$~}
\newcommand{\mun}{$\mu_{N}^2$~}
\newcommand{\ee}{$(e,e')$~}
\begin{document}
\title{Magnetic excitations in nuclei with neutron excess}
\author{G. Co', V. De Donno} 
\affiliation{Dipartimento di Fisica,
Universit\`a del Salento and INFN Sezione di Lecce, Via Arnesano,
I-73100 Lecce, ITALY} 
\author{M. Anguiano, A. M. Lallena}
\affiliation{Departamento de F\'\i sica At\'omica, Molecular y
Nuclear, Universidad de Granada, E-18071 Granada, SPAIN} 
\date{\today}

\bigskip
\begin{abstract}
The excitation of the $1^+$, $2^-$ and $3^+$ modes in $^{16}$O,
$^{22}$O, $^{24}$O, $^{28}$O, $^{40}$Ca, $^{48}$Ca, $^{52}$Ca and
$^{60}$Ca nuclei is studied with self-consistent random phase
approximation calculations. Finite-range interactions of Gogny type,
containing also tensor-isospin terms, are used.  We analyze the
evolution of the magnetic resonances with the increasing number of
neutrons, the relevance of collective effects, the need of a correct
treatment of the continuum and the role of the tensor force. 
\end{abstract}
\bigskip
\bigskip
\bigskip

\maketitle
\section{Introduction}
\label{sec:intro} 
The possibility offered by the new radioactive ion beam facilities to
produce nuclei with neutron excess opens new perspectives in the study
of nuclear excitations. Recently, we have investigated the electric,
natural parity, excitations of these nuclei by using a self-consistent
continuum Random Phase Approximation (CRPA) approach
\cite{don11a,don11b}.  In self-consistent approaches, the single
particle (s.p.) wave functions and energies are obtained by solving
the Hartree-Fock (HF) equations with the same effective interaction
used in the RPA calculations.  The values of the parameters of these
interactions are chosen to reproduce some ground state properties of a
large number of nuclei.  These fits produce universal
parameterizations of the force to be used for all nuclei, even for
those not yet explored by the experiment.
Self-consistent RPA approaches have greater prediction power than
their phenomenological counterparts, but they require a higher level
of accuracy in the calculations. For example, the dimension of the
s.p. configuration space, beyond a certain size, is not a problem in
phenomenological approaches since the effects of the truncation of the
s.p. basis are taken into account by changing the values of the
interaction parameters. This procedure cannot be used in
self-consistent approaches, because the interaction parameters are
chosen once forever. This drawback of the self-consistent RPA approach
is avoided if the full s.p. configuration space is used in the
calculation. This implies a proper treatment of the continuum part of
the s.p. spectrum.
In this work we present the results of our study of magnetic,
unnatural parity, excitations, conducted with our self-consistent CRPA
approach.  Since magnetic excitations are modes where nucleons with
different spin orientations vibrate ones against the other ones, it is
obvious that the restoring force is related to the spin-dependent
terms of the nuclear interaction. Microscopically, the most important
term of this part of the interaction is generated by one-pion
exchange, which is the longest range term of the interaction, and has
a tensor spin-isospin dependent component. This means that a realistic
description of magnetic excitations requires an effective interaction
which has both finite-range and tensor components. The CRPA approach
we have developed allows us to consider both these characteristics
without any approximation.

The role of the tensor term of the effective interaction has
been widely studied by using Skyrme interactions to calculate both
ground \cite{bro06,les07,col07,bri07,col08,ben09,mor10} and excited states
properties \cite{bai09a,bai09b,cao09,bai10} of various nuclei. However, in
this case the tensor term has a zero-range character.

We carried out our calculations by using finite range forces of Gogny
type which were first introduced in Ref. \cite{dec80}. Later, a
parameterization, named D1S, was chosen to reproduce binding energies
and surface properties of a large variety of nuclei
\cite{ber89,ber91}. In addition, neutron matter properties were
considered in the fit of the D1N parameterization
\cite{cha07t,cha08}. More recently \cite{gor09}, the D1M
parametrization was adjusted to reproduce, together with neutron
matter properties, a large number of binding energies and root mean
square charge radii within Hartree-Fock-Bogoliubov theory.  In our
work, together with the D1S  and D1M 
interactions, we also used two new parametrizations, called D1ST and
D1MT, which we have recently constructed by adding a finite-range
tensor-isospin term to the former two parameterizations \cite{ang11}.

We have conducted our investigation in various oxygen and calcium
isotopes where the s.p. levels below the Fermi surface are fully
occupied, and those above are empty. These isotopes are spherical.
Furthermore, we have verified the relevance of the pairing by doing
BCS calculations. We found presence of pairing effects only in 
$^{22}$O and $^{52}$Ca nuclei, where they are, however, so small, few
parts on a thousand in binding energies and root mean squared radii,
that we neglected them in our study.

In this presentation we focus our attention on the following points:
\begin{enumerate}
\item the evolution of the strength distribution of a specific
  magnetic multipole excitation with increasing neutron number;
\item the relevance of collective effects;
\item the need of a correct description of the continuum, and
\item the role of the tensor force. 
\end{enumerate}

In Sec. \ref{sec:form} we briefly recall the main features of our CRPA
approach and we give the basic expressions of the observables
calculated.  In Sec. \ref{sec:details} we describe the interactions
and the various types of calculations we have used in our
investigation.  In Sec. \ref{sec:results} we present a selected set of
the results we have obtained. We first discuss the excitation of the
$1^+$ mode, then we consider the $2^-$ and $3^+$ excitations.  We
summarize our main results and we draw our conclusions in
Sec. \ref{sec:conclusions}.

%-----------------------------------------------------

\section{Formalism}
\label{sec:form}
In this section we present the basic ideas of the method we use to
solve the CRPA equations. A detailed presentation can be found in
Ref. \cite{don11a}.

The starting point of the CRPA theory is the expression of the
operator that applied to the ground state generates the excited state
$|\nu\rangle$:
\beq 
Q^\dag_\nu \, = \, \sum_{ph} \, \sumint_{\epsilon_p} \,
\left[X^\nu_{ph}(\epsilon_p)\, a^\dag_p(\epsilon_p)\, a_h \, - \,
Y^\nu_{ph}(\epsilon_p)\, a^\dag_h \, a_p(\epsilon_p) \right] \, ,
\label{eq:qnu}
\eeq
where we have indicated with $a^\dag$ and $a$ the usual particle
creation and annihilation operators and with $X^\nu_{ph}$ and
$Y^\nu_{ph}$ the RPA amplitudes. In the above equation we have
explicitly indicated the dependence upon the variable $\epsilon_p$,
the energy of the particle state. We have indicated with the label $p$ the 
orbital and total angular momentum quantum numbers. 
The symbol $\displaystyle \sumint $ indicates a sum
on the discrete values of $\epsilon_p$ and an integration on the
continuous ones. The symbol $h$ indicates all the quantum numbers
characterizing a state below the Fermi surface, a hole state,
including its energy, which assumes discrete values only.

The CRPA secular equations whose solution provides the
values of $X$ and $Y$ can be written as
\beqn 
(\epsilon_p-\epsilon_h-\omega)\, X_{ph}^\nu(\epsilon_p)\,+
\nonumber    && \\  & &
\hspace*{-3.5cm} \sum_{p'h'} \, \sumint_{\epsilon_{p'}} \,
\left[v^J_{ph,p'h'}(\epsilon_p,\epsilon_{p'})\,
X_{p'h'}^\nu(\epsilon_{p'})\, + \,
u^J_{ph,p'h'}(\epsilon_p,\epsilon_{p'}) \, Y_{p'h'}^\nu(\epsilon_{p'})
\right] \, = \, 0 \,,
\label{eq:rpa1}
\\
(\epsilon_p-\epsilon_h+\omega) \, Y_{ph}^{\nu}(\epsilon_{p})\,+
\nonumber  && \\ & &
\hspace*{-3.5cm} \sum_{p'h'} \, \sumint_{\epsilon_{p'}}\,  \left[
v^{J*}_{ph,p'h'}(\epsilon_p,\epsilon_{p'})\,
Y_{p'h'}^{\nu}(\epsilon_{p}) \, + \,
u^{J*}_{ph,p'h'}(\epsilon_p,\epsilon_{p'}) \,
X_{p'h'}^{\nu}(\epsilon_{p'}) \right] \, = \, 0 \,.
\label{eq:rpa2}
\eeqn
In the above equations, $\omega$ labels the excitation energy and
the interaction terms have been defined as
\beqn 
v^J_{ph,p'h'}(\epsilon_p,\epsilon_{p'}) &=&  v^{J,{\rm
dir}}_{ph,p'h'}(\epsilon_p,\epsilon_{p'}) \, - \,   v^{J,{\rm
exc}}_{ph,p'h'}(\epsilon_p,\epsilon_{p'})  \, ,
\label{eq:v}
\eeqn and  \beqn u^J_{ph,p'h'}(\epsilon_p,\epsilon_{p'}) &=&
(-1)^{j_{p'} + j_{h'} - J} \,  v^J_{ph,h'p'}(\epsilon_p,\epsilon_{p'})
\, ,
\label{eq:u}
\eeqn
with
\beqn  && \hspace*{-1cm} v^{J,{\rm
dir}}_{ph,p'h'}(\epsilon_p,\epsilon_{p'}) \, = \, 
\sum_\alpha \,
\int d^3 r_1\,
\int d^3 r_2 \, \phi^*_p(\br_1,\epsilon_p) \,
\phi^*_{h'}(\br_2) \, V_\alpha(\br_1,\br_2) \, 
\phi_h(\br_1) \, \phi_{p'}(\br_2,\epsilon_{p'}) \, ,
\label{eq:vdir} \\ 
&& \hspace*{-1cm} v^{J,{\rm exc}}_{ph,p'h'}(\epsilon_p,\epsilon_{p'})
\, =\,  \sum_\alpha \,
\int d^3 r_1\, \int d^3r_2 \, 
\phi^*_p(\br_1,\epsilon_p)\, \phi^*_{h'}(\br_2) \, 
V_\alpha(\br_1,\br_2) \,  
\phi_{p'}(\br_1,\epsilon_{p'}) \,  \phi_h(\br_2) \,
\label{eq:vexc}
\eeqn
Here we have indicated with $\phi$ the s.p. wave
function. 

In our calculations we consider a two-body nucleon-nucleon  
interaction composed by terms of the form
\beq V_\alpha (\br_i,\br_j)= v_\alpha(|\br_i-\br_j|)\; O^\alpha_{i,j}
\, , \,\,\, %{\rm with} \,\, 
  \alpha=1,2,\ldots,6  \, ,
\label{eq:force1} 
\eeq
where $v_\alpha$ is a scalar functions of the distance between the two
interacting nucleons and $O^\alpha$ indicates the type of operator
dependence. Specifically we have considered the following six  
expressions:
\beq O^{\alpha}_{i,j} : 1\,,\,\,\btau(i)\cdot\btau(j)\,,\,\,
\bsigma(i)\cdot\bsigma(j)\,,\,\,
\bsigma(i)\cdot\bsigma(j)\,\btau(i)\cdot\btau(j)\,, 
%\nonumber \\ &~&
S(i,j)\,,\,\, S(i,j) \btau(i)\cdot\btau(j) 
\label{eq:fcahnnels}
\, .  \eeq
In the above expressions we have indicated with  $\bsigma$ the Pauli
matrix operator acting on the spin variable, with $\btau$ the
analogous operator for the isospin, and with
\beq S(i,j)\, = \,3 \, \frac {[\bsigma(i)\cdot (\br_i-\br_j)] \,
[\bsigma(j)\cdot (\br_i-\br_j)] }  {(\br_i-\br_j)^2}\, - \,
\bsigma(i)\cdot\bsigma(j) \eeq
the usual tensor operator. In the HF calculations we have implemented
this interaction with a density dependent zero-range spin-orbit term
as it is commonly done in the formulation of the Gogny interaction
\cite{dec80}. In our calculations the Coulomb and spin-orbit terms of
the interaction are considered in HF, but neglected in RPA
calculations. This breaks the complete self-consistency of our
calculations, however, these two terms of the interaction produce
small effects. Calculations done with Skyrme interactions indicates
that the effects of these two terms of the effective interaction, have
the tendency of canceling with each other \cite{sil06}. Results
obtained with Gogny interaction in medium-heavy nuclei indicate
noticeable effects in low-lying quadrupole and octupole excitations 
\cite{per05}. The role played by spin-orbit and Coulomb interactions in
RPA calculations is a topic which deserves further investigation.

The first step of our method of solving the CRPA equations
(\ref{eq:rpa1}) and (\ref{eq:rpa2}) consists in reformulating them in
terms of new unknown functions, called channel functions,  
which do not have an explicit
dependence on the continuous s.p. energy $\epsilon_p$:
\beq  f^\nu_{ph}(r) \, = \, \sumint_{\epsilon_p} \,
X^\nu_{ph}(\epsilon_p) \, R_p(r,\epsilon_p) \, ,
\label{eq:f}
\eeq
and
\beq  g^\nu_{ph}(r) \, = \, \sumint_{\epsilon_p} \,
Y^{\nu*}_{ph}(\epsilon_p) \, R_p(r,\epsilon_p)  \, .
\label{eq:g}
\eeq
In the above equations, we have indicated with $R$ the radial part of
the s.p. wavefunction. 

In this new reformulation of the CRPA equations, a set of algebraic
equations with unknowns depending on the continuous variable
$\epsilon_p$ has been changed into a set of integro-differential
equations with unknowns depending on the distance from the center of
coordinates. We solve this new system of equations by expanding the
channel functions $f$ and $g$ on a basis of Sturmian functions. 

The CRPA equations are solved by imposing that the particle is emitted
with specific values of energy and of orbital and total angular
momenta. These quantum numbers characterize, together with the quantum
numbers identifying the hole state, the so-called elastic channel
$p_0h_0$.  For a given value of $\omega$, the CRPA equations are then
solved for every elastic channel $p_0h_0$ allowed by the energy
conservation. In other words, the number of elastic channels is that
of the $ph$ pairs where the particle is in the continuum.

The solution of the CRPA equations provides the channel functions
$f^{p_0h_0}_{ph}(r)$ and $g^{p_0h_0}_{ph}(r)$ that allow us to
calculate the transition matrix elements induced by an operator $T_J$.
If this operator is of one-body type, it can be expressed as
\beqn
T_{JM}(\br)\, = \, \sum_{i=1}^A \, F_J(r_i) \, \theta_{JM}(\Omega_i)
\, \delta(\br_i-\br) \, , 
\label{eq:obop}
\eeqn 
where we have separated the dependence on the radial and on the
angular variables.  By using the above expression we can express the
transition matrix element as
\beqn \langle J \| T_J \| 0 \rangle_{p_0h_0} &=&  \,  \sum_{p h}\,
\left[ \langle j_p \| \theta_J \| j_h \rangle \,   \int {\rm d}r \,
r^2 \, (f^{p_0h_0}_{ph}(r))^* \, F_J(r) \, R_h(r)   \right.\nonumber
\\ && \hspace*{1.2cm} \left. + \, (-1)^{J+j_p -j_h} \,  \langle j_h \|
\theta_J \| j_p \rangle \,   \int {\rm d}r \, r^2 \,  R^*_h(r) \,
F_J(r) \, g^{p_0h_0}_{ph}(r) \right]  \, .
\label{eq:transs1}
\eeqn
Here, the double bar indicates the reduced angular momentum matrix
elements as defined by the Wigner-Eckart theorem which we consider
with the phase convention of Ref. \cite{edm57}.

In the present paper, for the photon excitation of
unnatural parity, magnetic, states we use the following expression for
the operator $T_J$:  
\beq T_{JM} \, = \,\mu_N \sum_{i=1}^A \left[ g_s^{(i)}\bs_i
+\frac{2}{J+1}g_l^{(i)}\bl_i \right] \cdot \left[
\bN \br^J_i \bY_{JM}(\Omega_i) \right] 
\delta(\br_i-\br) \, , 
\label{eq:ophot}
\eeq
where $\bY_{JM}$ indicates the vector spherical harmonics
\cite{edm57}, $\mu_N=e\hbar/2mc$ is the nuclear magneton and $g_l$ and
$g_s$ are the gyromagnetic factors for orbital angular momentum and
spin ($g_l=1$ , $g_s=5.586$, for protons, and $g_l=0$, $g_s=-3.826$,
for neutrons). 

For a given excitation energy $\omega$ and magnetic transition 
${\cal  M}J$ 
we calculated the $B$-value as the incoherent sum over all the elastic
channels $p_0h_0$,
\beq  B({\cal M}J)\!\uparrow \, =   \sum_{p_0h_0} |\langle \omega,J \|
T_J \| 0\rangle_{p_0h_0}|^2  \, .
\label{eq:bv} 
\eeq
The explicit expression of the s.p. matrix elements of the operator 
(\ref{eq:ophot}) is well known in the literature, (see, for example, 
Eq. (B.82) of Ref. \cite{rin80}).  

Eqs. (\ref{eq:obop}) and (\ref{eq:transs1}) are very general, and we
used them to evaluate inelastic electron scattering cross sections.
In this case, we used the expressions of the convection and
magnetization currents and those of the corresponding matrix elements
given in Ref. \cite{ama93}.

\begin{table}[b]
\begin{center}
\begin{tabular}{ccccccc}
\hline
\hline
 nucleus  && s.p. state & D1S & D1ST & D1M & D1MT \\
\hline
\hline
$^{16}$O & & proton $(1p_{1/2})^{-1}$ & -12.53 & -12.48 & -11.94 & -11.84 \\   
$^{22}$O & & neutron $(1d_{5/2})^{-1}$ &  -6.61 &  -6.19 &  -6.38 &  -6.27\\   
$^{24}$O & & neutron $(2s_{1/2})^{-1}$ &  -4.17 &  -4.18 &  -4.11 &  -4.17\\
$^{28}$O & & neutron $(1d_{3/2})^{-1}$ &  -0.96 &  -0.94 &  -0.87 &  -0.72\\
\hline
$^{40}$Ca & & proton $(1d_{3/2})^{-1}$ & -9.26 &  -9.18 &  -8.86 &  -8.72\\   
$^{48}$Ca & & neutron $(1f_{7/2})^{-1}$ & -9.48 &  -9.09 &  -9.33 &  -9.27\\
$^{52}$Ca & & neutron $(2p_{3/2})^{-1}$ & -5.58 &  -5.40 &  -5.56 &  -5.49\\   
$^{60}$Ca & & neutron $(1f_{5/2})^{-1}$ & -3.05 &  -2.96 &  -3.29 &  -3.09\\
\hline \hline   
\end{tabular}
\end{center} 
\caption{\small The least bound s.p. levels, and their energies, in
  MeV, for the nuclei considered in the work. 
}
\label{tab:sep}
\end{table}

The formalism we have just presented allows us to solve the CRPA
equations for energies above the continuum threshold,
i.e. excitation energies larger than the s.p. energy of the least bound
s.p. state. We indicate in Table \ref{tab:sep} the least bound
s.p. levels and their energies for the nuclei we have
investigated. For nuclei with equal number of protons and neutrons,
the least bound level is that of protons, for the other nuclei is
a neutron level. For excitation energies below the continuum
threshold, i.e. whose values are smaller than the absolute values of
the energies listed in Table \ref{tab:sep}, continuum and discrete RPA
produce the same solution. This has been numerically verified by using
a Fourier-Bessel formalism to solve the CRPA equations
\cite{deh82t,deh82,co06b}. In our work, for excitation energies below the
continuum threshold, we used the results obtained in the 
discrete RPA approach \cite{suh07,don08t,co09b,don09}.

%%%%%%%%%%%%%%%%%%%%%%%%%%%%%%%%%%%%%%%%%%%%%%%%%%%%%%%%%%%%%%%
\section{Details of the calculations}
\label{sec:details}

Our calculations are based on two different parameterizations of the
Gogny interaction, the more traditional D1S force \cite{ber91} and the
new D1M force \cite{gor09} obtained from a fit to about 2000 nuclear
binding energies and 700 charge radii. The D1S and D1M forces describe
the empirical saturation point of symmetric nuclear matter and
reproduce rather well the behavior of the equations of state
calculated with microscopic approaches \cite{akm98,gan10}. The
situation for pure neutron matter is different, because the behavior
of the D1S equation of state at high densities is unphysical. The D1M
force produces a neutron equation of state which has a plausible
behavior at high densities, even though it does not reproduce the
results of modern microscopic calculations. In addition to these two
forces, we also used other two parameterizations of the Gogny force
containing a tensor-isospin term, the D1ST and D1MT interactions
\cite{ang11}. In mean-field calculations, the inclusion of the tensor
force does not modify the nuclear and neutron matter equations of
state.   
We have constructed these interactions by adding to the
corresponding Gogny parameterization a tensor-isospin term obtained by
multiplying the analogous term of the microscopic Argonne V18
interaction by a function which modifies its behavior at short
internucleonic distances. This function contains a single parameter
whose value determines the strength of the tensor force.  We have
chosen the value of this parameter to reproduce the experimental
energy of the first $0^-$ excited state and the splitting between the
s.p.  energies of the neutron $(1p_{3/2})^{-1}$ and $(1p_{1/2})^{-1}$
levels in the \oxy nucleus. In what follows we have used the
superscript $-1$ to indicate a hole s.p. state, i.e. below the Fermi
surface.

As last detail regarding our CRPA calculations, it is worth to mention
that we found convergence in our results when we included at least 10
expansion coefficients of the Sturm-Bessel basis used in the
calculations.

In this work, we have compared the results obtained with discrete
RPA (DRPA) calculations with those found in CRPA ones. We have
described in the previous section how we solve the CRPA equations.
The solution of the DRPA equations is based on the expansion of the HF
s.p. wave functions on a harmonic oscillator basis and a subsequent
diagonalization of the secular equations expressed in a matrix
form. The expansion on the harmonic oscillator basis imposes an
exponentially decaying asymptotic behavior also to the s.p. wave
functions with positive energy. This implies a discretization of the continuum
that requires a truncation of the s.p. space. In our calculations we
have obtained convergence in the results by using 50 harmonic
oscillator expansion coefficients and s.p. energies up to 150 MeV. 

In the following, we shall compare our RPA results with those of the
independent particle model (IPM) obtained by switching off the
residual interaction in RPA calculations. The IPM calculations have
been performed in both discrete and continuum cases, and we indicate
the corresponding results as DIPM and CIPM, respectively.

Our formalism is constructed to treat spherical systems.  For this
reason, we have chosen to study four oxygen isotopes, $^{16}$O,
$^{22}$O, $^{24}$O and $^{28}$O, and four calcium isotopes,
$^{40}$Ca, $^{48}$Ca, $^{52}$Ca and $^{60}$Ca, where the s.p. levels
below the Fermi surface are fully occupied. In these nuclei there are
not deformations and the pairing effects are negligible as indicated
by the results of the deformed Hartree-Fock-Bogoliubov calculations
of Ref. \cite{del10}.  Recently, we have studied the ground state
properties of these isotopes, and also those of other heavier nuclei,
with HF calculations done with finite and zero range forces and with
relativistic Hartree calculations \cite{co11b}. We have found a good
convergence of the results of these three types of calculations in all
the properties studied. 

%----------------------------------------------------------------
\section{Results}
\label{sec:results}

In this section we present a selection of the results we have obtained
in our study.  The presentation is organized as follows.  We first
discuss, with some detail, the $1^+$ excitation, especially the
results obtained for the oxygen isotopes. In a following subsection
we present results concerning the $2^-$ and $3^+$ magnetic
excitations. 

We have carried on our calculations by using the four interactions
introduced in the previous section.  We have observed that the results
obtained with the D1S and D1M forces are similar, as well as those
obtained with the D1ST and D1MT interactions. For this reason, in the
following, we shall present mainly the results obtained with the D1S
and D1ST forces, eventually quoting those obtained with the other two
interactions when this is relevant for the discussion.

\subsection{Magnetic dipole response}
\label{sec:1+} 

\begin{figure}[ht]
\begin{center}
\includegraphics [scale=0.6]{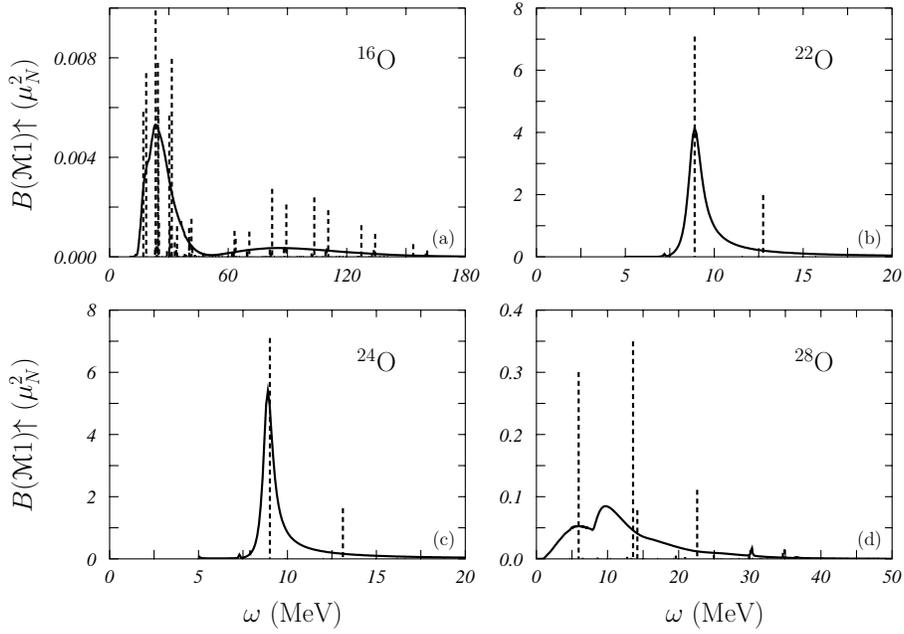}
\caption{\small \bmon values obtained with the D1S interaction for 
  the oxygen isotopes under investigation 
  as a function of the excitation energy. The 
  full lines show the CRPA results, while the vertical 
  dashed lines indicate the DRPA results. 
}
\label{fig:dox1+}
\end{center}
\end{figure}

In Fig. \ref{fig:dox1+} we compare the \bmon results for the four
oxygen isotopes under investigation obtained with the D1S interaction
in DRPA (dashed vertical lines) and CRPA (full lines) calculations.
In order to show a palusible comparison between the reusults of the
two types of calculations we have investigated rather different energy
range for the four isotopes. We remark that also the value of the
strengths we have obtained is rather different for the four isotopes. 

%%%%%%%%%%%%%%%%%%%%%%%%%%%%%%%%%%%%%%%%%%%%%%%%%%%%%%%%%%%%%%%%%%%%%%
% oxy IPM 1+ 
%%%%%%%%%%%%%%%%%%%%%%%%%%%%%%%%%%%%%%%%%%%%%%%%%%%%%%%%%%%%%%%%%%%%%%
\begin{figure}[ht]
\begin{center}
\includegraphics [scale=0.6] {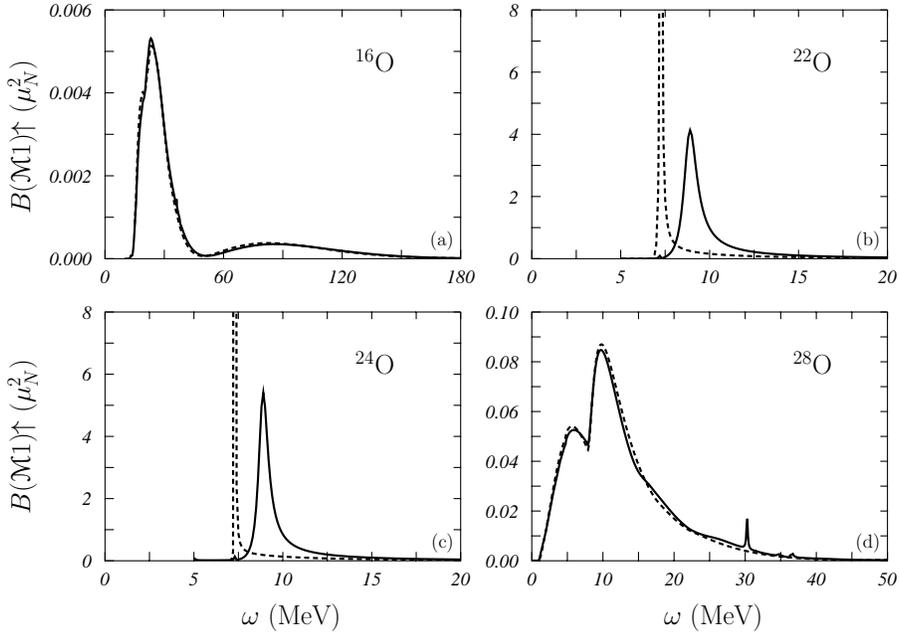}
\caption{\small \bmon values obtained with the D1S interaction for 
  the oxygen isotopes under investigation 
  as a function of the excitation energy.
  The full lines show the CRPA results, while the 
  dashed curves the CIPM results. 
}
\label{fig:ipmox1+}
\end{center}
\end{figure}

The positions of the peaks of the DRPA results correspond to those of
the continuum responses. It is remarkable the agreement between the
main peaks of the discrete and continuum responses in $^{22}$O and
$^{24}$O.  Only the results for the $^{28}$O nucleus show a very broad
continuum response that the discrete calculations can hardly
reproduce. 

A feature common to all our results is already evident in
Fig. \ref{fig:dox1+}.  The \bmon strengths in the $^{16}$O and
$^{28}$O nuclei are orders of magnitude smaller than those of the
other two isotopes. In our model, the $1^+$ excitation in $^{16}$O and
$^{28}$O is generated by $2 \hbar \omega$ particle-hole (p-h)
configurations, since, in the ground states of these nuclei, the
nucleons fully occupy all the spin-orbit partner levels below the
Fermi surface. On the contrary, in $^{22}$O and $^{24}$O the neutron
$(1d_{3/2})$ level is empty, while the $(1d_{5/2})^{-1}$ level is
occupied. In this last case, since a $1^+$ transition between these
two states, a $0 \hbar \omega$ transition, is allowed, the
corresponding \bmon strengths are much larger than those of $^{16}$O
and $^{28}$O.

The role of the residual interaction on the energy distribution of the
\bmon strengths is shown in Fig. \ref{fig:ipmox1+} where we compare
the results obtained with the D1S interaction in CIPM and CRPA
calculations, shown by the dashed and full lines respectively.  We
observe an almost exact overlap of the two results in $^{16}$O and
$^{28}$O nuclei. The situation is more interesting for the other two
oxygen isotopes where the effective interaction pushes the peak
position at higher energies and, at the same time, spreads the
strength.

\begin{table}[t]
\begin{center}
\begin{tabular}{cccccc}
\hline\hline
            &  &   D1S    &    D1ST   &    D1M    &    D1MT  \\
\hline
\hline
  $^{22}$O  &DIPM  & 8.796 & 8.866 & 8.803 & 8.814 \\  
            &DTDA  & 8.796 & 8.866 & 8.803 & 8.814 \\ 
            &DRPA  & 7.618 & 7.155 & 7.887 & 7.581 \\
            &CIPM  & 8.106 & 7.890 & 8.106 & 7.927 \\
            &CRPA  & 6.225 & 5.599 & 6.385 & 5.906 \\
\hline
  $^{24}$O  &DIPM  & 8.438 & 8.425 & 8.435 & 8.432 \\ 
            &DTDA  & 8.438 & 8.425 & 8.435 & 8.432 \\ 
            &DRPA  & 7.300 & 6.782 & 7.548 & 7.242 \\
            &CIPM  & 8.435 & 8.437 & 9.168 & 8.478 \\
            &CRPA  & 6.285 & 6.643 & 6.497 & 6.400 \\
\hline \hline
\end{tabular}
\end{center} 
\caption{\small 
  Total \bmon strengths 
  in $\mu^2_N$ MeV units, obtained in discrete and 
  continuum calculations with various interactions for the 
  $^{22}$O and $^{24}$O nuclei. 
}
\label{tab:bmcox1+}
\end{table}

Further information on the role of the residual interaction is given
by the total \bmon \, strengths of the $^{22}$O and $^{24}$O nuclei shown
in Table \ref{tab:bmcox1+} where we compare the results obtained with
all the four interactions introduced in Sec. \ref{sec:details}.  The
total strengths of the discrete calculations have been obtained by
summing all the \bmon values found in the diagonalization
procedure. In the continuum cases, the values given in the table have
been obtained by integrating the $^{22}$O and $^{24}$O strength
distributions shown in the panels of Figs. \ref{fig:dox1+} and
\ref{fig:ipmox1+}, i.e. up to a maximum energy of 20 MeV.  In the table we
also show the results obtained in discrete Tamm-Dancoff calculations
(DTDA) performed by switching off the RPA terms related to the $Y$
amplitudes. 

The results shown in the table indicate that, independently from the
interaction, the values of the total strengths of the DIPM and DTDA
calculations are identical.  The residual interaction in TDA
calculations redistributes the IPM strength between the various
excited states, which appear at different excitation energies, without
modifying its total value.  The situation changes when ground state
correlations are considered.  DRPA results show that the values of the
total strengths are reduced with respect to those of the DIPM.  This
effect of the ground state correlations is evident also in the results
of the continuum calculations where we
found even larger reduction factors. 

%%%%%%%%%%%%%%%%%%%%%%%%%%%%%%%%%%%%%%%%%%%%%%%%%%%%%%%%%%%%%%%%%%%%%%
% 22O (e,e) a
%%%%%%%%%%%%%%%%%%%%%%%%%%%%%%%%%%%%%%%%%%%%%%%%%%%%%%%%%%%%%%%%%%%%%%
\begin{figure}[ht]
\begin{center}
\includegraphics [scale=0.8]{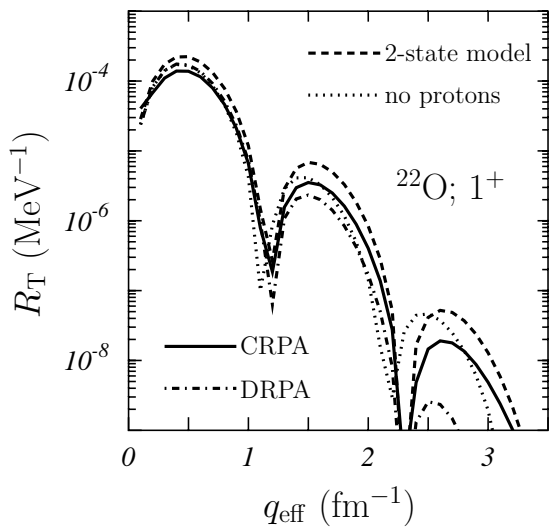}
\caption{\small Transverse response of inelastic electron
scattering $1^+$ excitation in $^{22}$O
at the peak energy as a function of the effective momentum transfer. 
The DRPA and CRPA results obtained with the 
D1S force are compared. The dashed curve has been obtained by 
considering only the two main neutron p-h excitations in a DRPA 
calculation. The dotted curve corresponds to a DRPA calculation 
where all the proton excitations have been eliminated.
}
\label{fig:ee22a}
\end{center}
\end{figure}

We show in Fig. \ref{fig:ee22a} the transverse responses of
the $^{22}$O nucleus for an inclusive electron scattering process
calculated at the peak energy with different models. 
The responses are shown as a function of the effective momentum 
transfer \cite{hei83}. 
In these
calculations we used the D1S interaction.  The result of the CRPA
calculation has been obtained by integrating the responses on the
excitation energy below the peak.  The integration limits have been
chosen such as the result is numerically stable.
 
The results shown in the figure indicate that the differences between
continuum and discrete RPA results arise mainly at high momentum
values. In order to understand the source of these differences we made
a DRPA calculation by using only the two main neutron p-h excitations:
$[(1d_{3/2})(1d_{5/2})^{-1}]$ and $[(2d_{3/2})(1d_{5/2})^{-1}]$.  The
result of this calculation is shown by the dashed line, which has a
rather different behavior with respect to the CRPA response for all
the momentum transfer values.  We performed another calculation by
using DRPA wave functions where all the proton contributions have been
eliminated.  The result of this calculation is shown by the dotted
line.  In this case we observe a good agreement with the CRPA result
below 2 fm$^{-1}$, and also the peak at about 2.7 fm$^{-1}$ is
reproduced in its gross features.  This indicates that in CRPA the
role of the main neutron s.p. components is enhanced with respect to
DRPA. The proton contributions in DRPA affect the response at high
momentum transfer since they are produced by $2\hbar\omega$
excitations that have higher Fourier components than the main neutron
configurations, which are $0\hbar\omega$ excitations. In DRPA
calculations, these proton contributions generate a destructive
interference with those of the neutrons and lower the transverse
response at large momentum transfer. A similar study in $^{24}$O
presents analogous features.

After discussing the role of the continuum and that of the residual
interaction, we now analyze the effects of the tensor part of the
residual interaction. In Table \ref{tab:bmcox1+} we compare the total
\bmon strengths obtained with the tensor force, the D1ST and D1MT
results, with those obtained without it, the D1S and D1M results.  The
effect of the tensor on the IPM strengths is negligible. More
interesting are the effects on the RPA results where the inclusion of
the tensor force lowers the values of the total \bmon strengths in
almost all the cases we have considered, the only exception being the
D1S results in $^{24}$O obtained in CRPA calculations. 
This last effect is probably due to the
truncation of the integration at 20 MeV, since we observe that around
this energy the response obtained with the tensor force drops more
quickly than the other one.

%%%%%%%%%%%%%%%%%%%%%%%%%%%%%%%%%%%%%%%%%%%%%%%%%%%%%%%%%%%%%%%%%%%%%%
% oxy 1+ tens 
%%%%%%%%%%%%%%%%%%%%%%%%%%%%%%%%%%%%%%%%%%%%%%%%%%%%%%%%%%%%%%%%%%%%%%
\begin{figure}[ht]
\begin{center}
\includegraphics [scale=0.6]{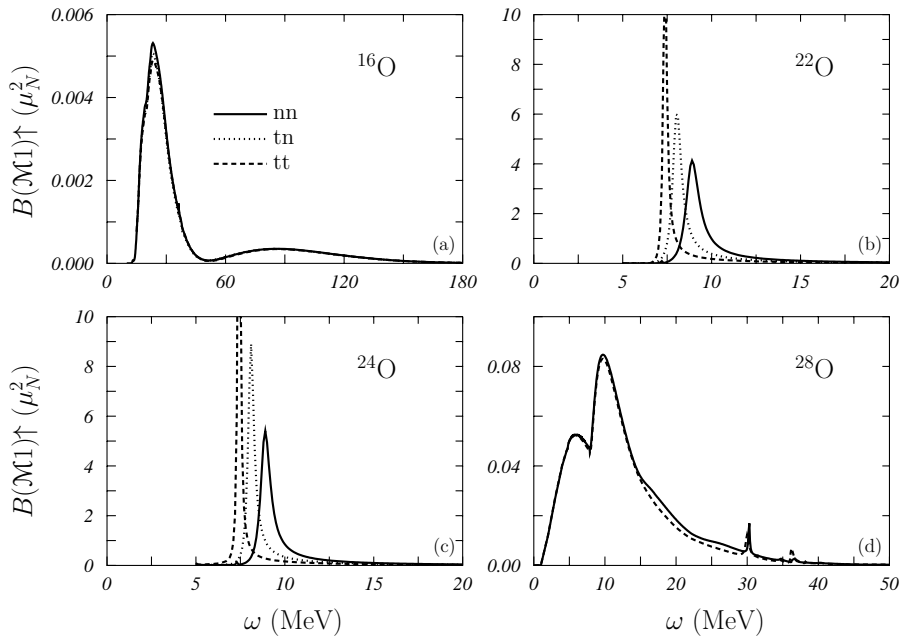}
\caption{\small  The CRPA results for the 
 \bmon strengths of the oxygen isotopes under investigation 
 as a function of the excitation energy. 
 The full lines, already shown in Fig. \ref{fig:dox1+} and
 \ref{fig:ipmox1+}, and here labelled nn, 
 indicate the results obtained with the D1S interaction in both HF and CRPA
 calculations.  The dotted lines, labelled tn,
 have been obtained by using the D1ST interaction in the HF calculation 
 and the D1S force in the CRPA ones. 
 The dashed lines, labelled tt show the results
 obtained by using the D1ST interaction in both HF and CRPA
 calculations.  
 }
\label{fig:tensox1+}
\end{center}
\end{figure}

The effects of the tensor force on the energy distributions of the
\bmon values are shown in Fig. \ref{fig:tensox1+}, where we
show CRPA results only and, following the nomenclature of
Ref. \cite{ang11}, we indicate with nn the results of calculations
done without tensor force, with tn those where the tensor force is
used only in HF calculations and not in RPA, and with tt the results
obtained by using the tensor force in both HF and RPA calculations.

The results of these three calculations almost overlap for the
$^{16}$O and $^{28}$O nuclei. We have observed that already the
results of Fig. \ref{fig:ipmox1+} indicated a small sensitivity to the
full residual interaction of the $1^+$ excitation in these two nuclei,
therefore, it is not surprising that the inclusion of the tensor force
does not change the situation.  In addition, these results confirm
that in HF calculations the tensor effects are irrelevant when all the
spin-orbit particle levels are occupied.

The situation changes for $^{22}$O and $^{24}$O. In these cases, the
effect of the tensor force consists in lowering the position of the
peaks of the response.  It is possible to identify two different
sources of this effect.  A first one is already present at the HF
level, as we can see by comparing nn and tn results.  The second
source is a genuine RPA effect, as we deduce by observing the tt
results. Our tensor terms are attractive in the RPA description of the
$1^+$ excitation.

The effect observed in the tn results is due to a change of the
s.p. neutron energies. In the case under investigation, the
s.p. energies of interest are those of the $(1d_{5/2})^{-1}$ and the
$(1d_{3/2})$ neutron levels. In our HF calculations, the inclusion of
the tensor term enhances the value of the energy of the first state
and lowers that of the second one, and this reduces the energy
difference.  Otsuka and collaborators \cite{ots05,ots06} pointed out
an effect of the tensor force which produces a lowering of the energy
differences between spin-orbit partner levels. This effect appears in
nuclei where not all the spin-orbit partner levels of a certain type
of nucleons (protons or neutrons) are occupied and affects the
s.p. energies of the nucleons of the other type. Therefore, in the two
nuclei under investigation, the proton s.p. levels should be affected,
as we have verified it happens in our HF calculations.  In
Ref.~\cite{ang11} we have investigated the presence, and the
consequences, of this effect, which we called Otsuka effect, in
various nuclei, including $^{22}$O and $^{24}$O. However, in the case
we are discussing now, the changes of the proton s.p. energies have
small effects on the main peak of the $1^+$ excitation in $^{22}$O and
$^{24}$O which is dominated by the neutron excitation indicated
above. This means that in our results we do not observe the effect
pointed out by Otsuka et al., but a similar one. In the present case,
the s.p. energies which are modified are those of the nucleons of the
same type of those where not all the spin-orbit partner levels are
occupied. 

%%%%%%%%%%%%%%%%%%%%%%%%%%%%%%%%%%%%%%%%%%%%%%%%%%%%%%%%%%%%%%%%%%%%%%
% 22O (e,e) b
%%%%%%%%%%%%%%%%%%%%%%%%%%%%%%%%%%%%%%%%%%%%%%%%%%%%%%%%%%%%%%%%%%%%%%
\begin{figure}[ht]
\begin{center}
\includegraphics [scale=0.8]{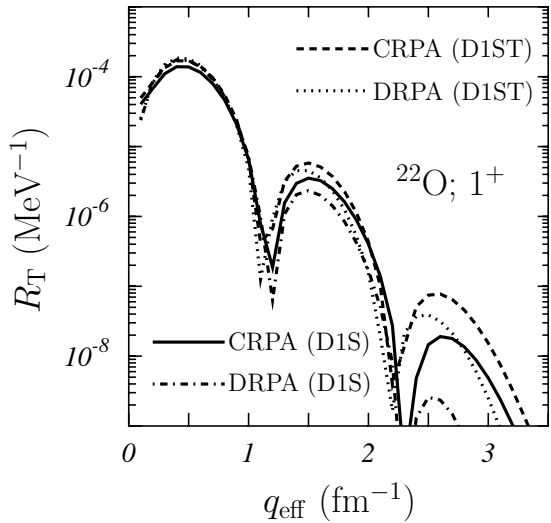}
\caption{\small 
Transverse response of the inelastic electron scattering 
$1^+$ excitation in the $^{22}$O nucleus at the peak energy as a
function of the effective momentum transfer.
We compare DRPA and CRPA results obtained with and
without tensor force.
}
\label{fig:ee22b}
\end{center}
\end{figure}

The study of the tensor effect has been done also by investigating the
electron scattering responses. We show in Fig. \ref{fig:ee22b} the
transverse $1^+$ responses of $^{22}$O calculated at the peak energy
with and without tensor force, in both discrete and continuum RPA
frameworks.  We observe that in both type of calculations the tensor
increases the responses at large $q$ values. The analysis of the
results of Fig. \ref{fig:ee22a} indicates that the peak at $q=2.7
\,\,{\rm fm}^{-1}$ is lowered by the presence of proton
excitations. Therefore the results of Fig. \ref{fig:ee22b} indicate
that the tensor force quenches the proton contribution, enhancing the
role of the main neutron excitation. We have done analogous
calculations in $^{24}$O and we have observed similar effects.

The study of the $1^+$ excitation we have just presented for the
oxygen isotopes has been carried on also for the four calcium isotopes
under investigation. The main features pointed out in the discussion
regarding the oxygen isotopes have been found also in this case. 
The \bmon values of the $^{40}$Ca and $^{60}$Ca isotopes are orders of
magnitude smaller than those obtained in $^{48}$Ca and $^{52}$Ca.  In
this case, the cause of the effect is the occupancy of the $1f$
spin-orbit partner s.p. levels.  They are both empty in $^{40}$Ca,
and occupied in $^{60}$Ca, while for the other two nuclei the
$(1f_{7/2})^{-1}$ level is occupied, while the $(1f_{5/2})$ level is
empty. The $1^+$ transition between these two levels is allowed and
this produces the increase of the strength of various order of
magnitude in $^{48}$Ca and $^{52}$Ca.  There is a small difference
with respect to the oxygen case. In $^{24}$O the $(2s_{1/2})^{-1}$ was
occupied, but this did not produce other $0 \hbar \omega$ $1^+$
excitations. In $^{52}$Ca the $(2p_{3/2})^{-1}$ is occupied, while its
spin-orbit partner, the $(2p_{1/2})$ level is empty. In this nucleus,
the transition between these two states adds another $0 \hbar \omega$
component to the $1^+$ excitation.

In analogy with the oxygen results, we have found that the total
strengths of the DIPM and DTDA calculations are conserved, while the
values of the total DRPA strengths are smaller than the two previous
ones. In this last case, the inclusion of the tensor force further
reduces the values of the total strengths, without exceptions.  The
same happens for CIPM and CRPA calculations.  The calculations in
calcium isotopes confirm that DRPA and CRPA produce peaks at the same
excitation energies. Also the general effect of the residual
interaction in RPA calculations has been confirmed. Finally, we have
observed that also for the calcium isotopes the residual interaction
moves the positions of the peaks at higher energies with respect to
those obtained in IPM calculations.

%%%%%%%%%%%%%%%%%%%%%%%%%%%%%%%%%%%%%%%%%%%%%%%%%%%%%%%%%%%%%%%%%%%%%%
% 1+n 
%%%%%%%%%%%%%%%%%%%%%%%%%%%%%%%%%%%%%%%%%%%%%%%%%%%%%%%%%%%%%%%%%%%%%%
\begin{figure}[ht]
\begin{center}
\includegraphics [scale=0.7]{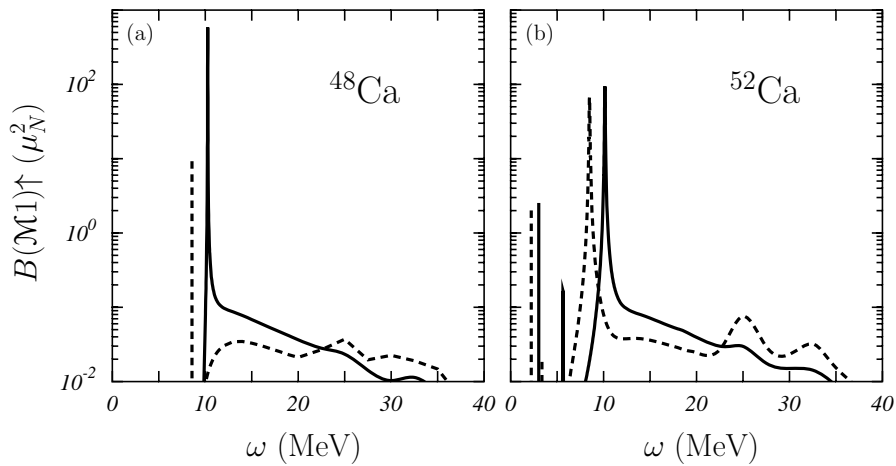}
\caption{\small 
Energy distribution of the \bmon strength for two calcium isotopes. 
Full and dashed curves indicate the
CRPA results obtained with D1S and D1ST interactions, respectively.
The strength below the continuum threshold has been obtained in a 
DRPA calculation. 
}
\label{fig:tensca1+}
\end{center}
\end{figure}

We show in Fig. \ref{fig:tensca1+} the energy distributions of the
\bmon strengths calculated with the D1S and D1ST interactions for the
$^{48}$Ca and $^{52}$Ca nuclei.  The strength below the continuum
  threshold has been obtained in a DRPA calculation.  The comparison
  between the curves shown in the figure illustrates the role of the
  tensor force.  We do not show the results for $^{40}$Ca and
$^{60}$Ca since, in this case, the effects of the tensor force are
negligible, as it happens in $^{16}$O and $^{28}$O nuclei.  We recall
that in these nuclei, all the spin-orbit partner levels are fully
occupied.  In Fig. \ref{fig:tensca1+} we used a logarithmic scale to
emphasize the widths of the peaks.  The role of the tensor force is
analogous to that pointed out in the discussion of
Fig. \ref{fig:tensox1+}.  The position of the peak is lowered when the
tensor force is considered.  In the case of the $^{48}$Ca nucleus, the
size of the effect is large enough to push the main resonance peak
below the particle emission threshold. We should remark that when the
D1M interaction is used, the peak energy is below the particle
emission threshold already in the calculation without tensor force.
Its inclusion pushes further down the position of the peak. 

The role of the tensor force is more clear in $^{52}$Ca where the main
excitation peak is always above threshold as it is shown in the panel
(b) of Fig. \ref{fig:tensca1+}. In this case, it is possible to
observe in a very clean way the lowering of the peak positions induced
by the tensor force.  The peaks below the particle emission
  threshold, obtained by a DRPA calculation, are dominated by the
  $[(2p_{1/2})(2p_{3/2})^{-1}]$ neutron transition.  The main peaks,
in the continuum, are instead dominated by the
$[(1f_{5/2})(1f_{7/2})^{-1}]$ neutron transition.  We have verified
that also in this case, as discussed in detail for the oxygen
isotopes, the global effect of the tensor force is produced by the sum
of an effect on the s.p. energies and a genuine effect in the CRPA
calculation.

Experimentally, the $1^+$ excitation in $^{48}$Ca has been studied by
using inelastic electron scattering \cite{ste80,ste83,knu83}. This
investigation has identified an isolated $1^+$ excitation at 10.23
MeV, with \bmon$= 4.0 \pm 0.4$ \mun. Our CRPA calculations generate the
main $1^+$ peak at 10.15 MeV, when the D1S force is used, and at 8.56
MeV,  therefore below the continuum threshold, when we consider
the D1ST force.  The corresponding \bmon values related to the main
peaks are, respectively, 9.72 and 9.27, in \mun units.  The values of
the total, integrated, \bmon strengths are 10.66 and 10.07 \mun MeV,
respectively, to be compared with the almost 12 \mun MeV predicted by
the IPM model.  The CRPA calculations done with the D1M and D1MT
interactions generate peaks at 9.25 and 8.58 MeV with \bmon values of
10.31 and 9.99 \mun, respectively.

%%%%%%%%%%%%%%%%%%%%%%%%%%%%%%%%%%%%%%%%%%%%%%%%%%%%%%%%%%%%%%%%%%%%%%
% 48Ca (e,e)
%%%%%%%%%%%%%%%%%%%%%%%%%%%%%%%%%%%%%%%%%%%%%%%%%%%%%%%%%%%%%%%%%%%%%%
\begin{figure}[ht]
\begin{center}
\includegraphics [scale=0.75]{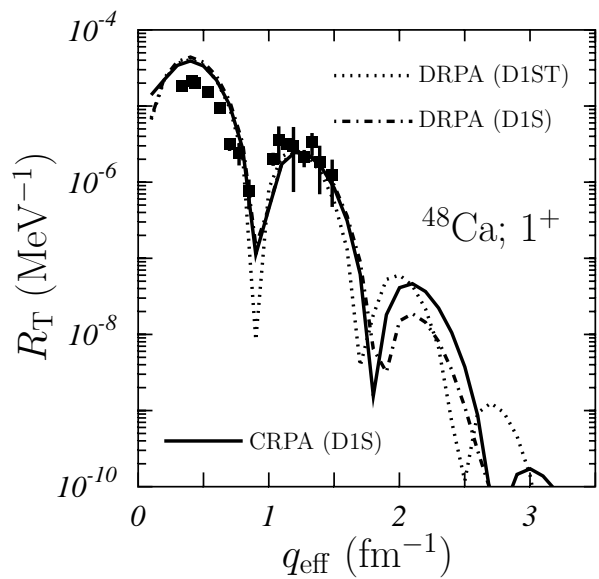}
\caption{\small 
Transverse responses of the inelastic electron scattering 
for the $1^+$ excitation in the $^{48}$Ca nucleus at the peak energy
as a function of the effective momentum transfer.
We show the DRPA and CRPA results obtained with the D1S interaction
and the DRPA result obtained with the D1ST interaction. 
The experimental data are from Ref. \cite{ste83}.
}
\label{fig:ee481+}
\end{center}
\end{figure}

The inclusive \ee transverse responses, obtained by using the RPA wave
functions in the peak energy, are compared in Fig. \ref{fig:ee481+}
with the experimental data of Ref. \cite{ste83}.  In this figure, we
compare the continuum and discrete results obtained with the D1S
interaction, since in both cases the peak is positioned above the
continuum threshold.  The dotted line indicates the result of a DRPA
calculation with the D1ST interaction which coincides with the CRPA
results since the excitation energy is below the continuum threshold.

We remark the agreement of the three calculations in the two peaks at
lower momentum values. In analogy to what we have observed in the case
of the $^{22}$O nucleus, the differences between the various results
arise at momentum transfer values larger than 2 fm$^{-1}$. These
differences are related to the proton excitations, as we have verified
by applying also in this case the same type of investigation done with the
results of Fig.~\ref{fig:ee22a}. We found, again, that the
contribution of these excitations is smaller in CRPA than in DRPA wave
functions. Also in the present case, the contibution of the proton
excitations is further reduced by the tensor force.

The comparison with the experimental data of Ref.~\cite{ste83}
indicate that our results overestimate the data of the first peak.  At
$q=0.4$ fm$^{-1}$ we need quenching factors of 0.48 and 0.51,
respectively, for the D1S and D1ST results. These values are similar
to those of the quenching factors necessary to reproduce the
experimental \bmon value. The application of a global quenching factor
to our responses would spoil the agreement with the data in the second
peak. Evidently, the physics behind the disagreement between
experimental and calculated responses is momentum transfer dependent,
and cannot be described by a single number, i. e. by using a quenching
factor.

\subsection{Magnetic excitations beyond the dipole} 
\label{sec:2-} 
In this section, we present our results for the magnetic excitations
with angular momentum larger than 1. We shall concentrate on the $2^-$
and $3^+$ excitation modes. These two modes are composed by p-h
excitations with intrinsic different characteristics.  The $2^-$ mode
is composed by p-h transitions between different major shells.  On the
contrary, in the $3^+$ mode p-h excitations within the same major
shell are allowed. We investigate whether this difference produces
remarkable effects on observables.

The study of \bmtw and \bmth total strengths indicates behaviours
similar to those shown by the \bmon strenghts of
Table~\ref{tab:bmcox1+}.  Also in the present cases, within the limits
of the numerical accuracy, the values of the DIPM and DTDA total
strengths coincide. In RPA calculations (both discrete and continuous)
the total \bmtw and \bmth values are smaller than those obtained in
DIPM calculations.  The only difference with respect to the \bmon
cases is related to the strenghts obtained by including the tensor
interaction which, in the \bmtw and \bmth cases, show an increase of a
few percent with respect to those obtained without tensor.

%%%%%%%%%%%%%%%%%%%%%%%%%%%%%%%%%%%%%%%%%%%%%%%%%%%%%%%%%%%%%%%%%%%%%%
% oxy Discrete 2- 
%%%%%%%%%%%%%%%%%%%%%%%%%%%%%%%%%%%%%%%%%%%%%%%%%%%%%%%%%%%%%%%%%%%%%%
\begin{figure}[ht]
\begin{center}
\includegraphics [scale=0.6]{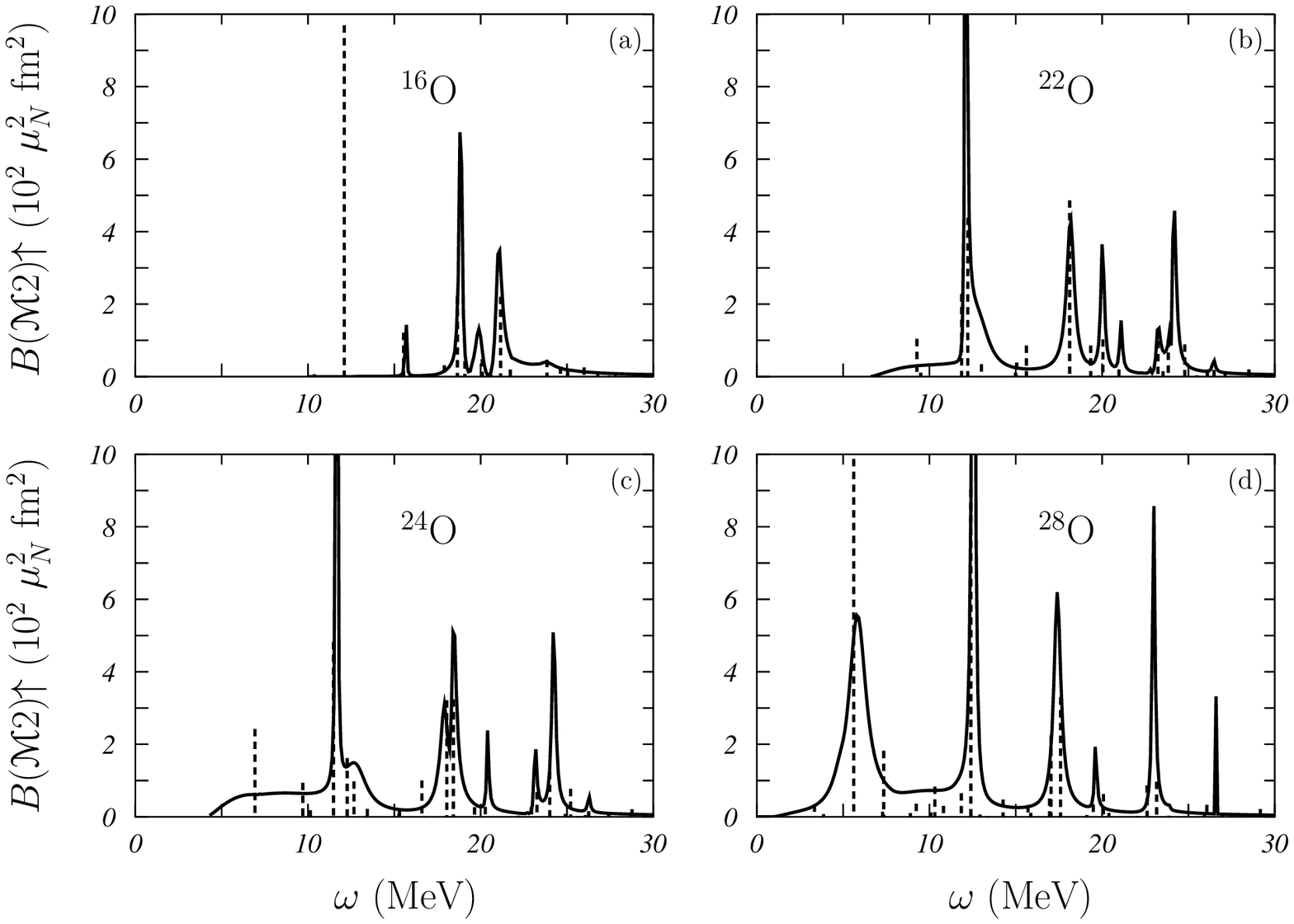}
\caption{\small  
The same as Fig.~\ref{fig:dox1+} for the \bmtw values. 
}
\label{fig:dox2-}
\end{center}
\end{figure}

We make now a more detailed analysis of the strength distributions of
these excitation modes.  We compare in Fig.~\ref{fig:dox2-} the
results obtained in DRPA and CRPA calculations for the
$2^-$ excitation in the oxygen isotopes, when the D1S interaction is
used. For all the isotopes considered we observe a good agreement
between the position of the peaks in discrete and continuum results.
In $^{16}$O the DRPA calculation produces a sharp peak at 12.10 MeV,
which is below the continuum threshold. This peak is dominated by the
s.p. transitions $[(1d_{5/2})(1p_{1/2})^{-1}]$ of both protons and
neutrons, whose unperturbed excitation energies are 10.27 MeV and
10.05 MeV respectively.  The figure shows that the size of the \bmtw
strength for the $^{16}$O and $^{28}$O is comparable with that of the
other two isotopes, contrary to what we found for the $1^+$
excitation.

%%%%%%%%%%%%%%%%%%%%%%%%%%%%%%%%%%%%%%%%%%%%%%%%%%%%%%%%%%%%%%%%%%%%%%
% oxy Discrete 2- 
%%%%%%%%%%%%%%%%%%%%%%%%%%%%%%%%%%%%%%%%%%%%%%%%%%%%%%%%%%%%%%%%%%%%%%
\begin{figure}[ht]
\begin{center}
\includegraphics [scale=0.6]{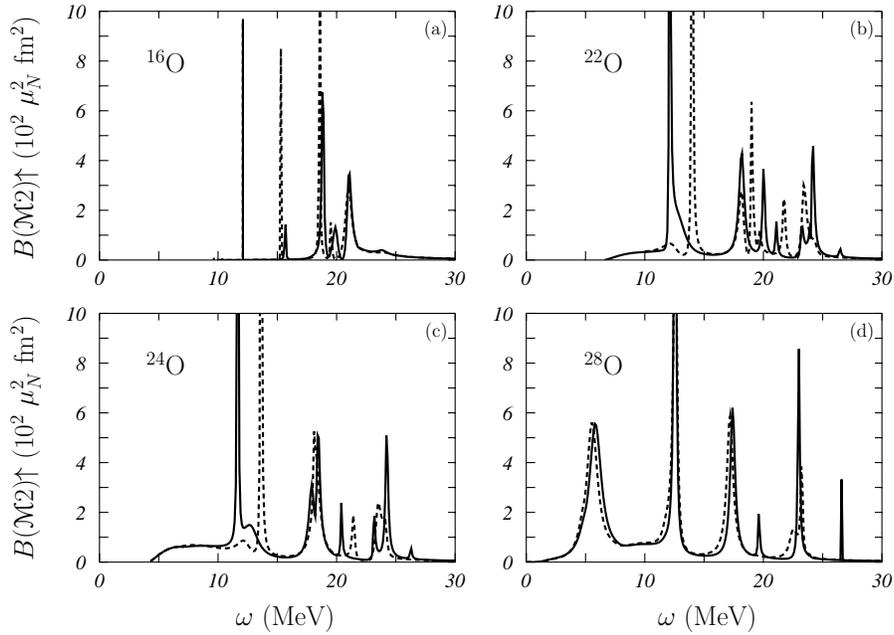}
\caption{\small \bmtw values for the four oxygen isotopes under
  investigation as a function of the excitation energy. 
  The full and the dashed curves show the CRPA results 
  obtained, respectively, with the D1S and D1ST interactions.}
\label{fig:tensox2-}
\end{center}
\end{figure}

The effect of the tensor force is shown in Fig.~\ref{fig:tensox2-}
where the CRPA results obtained with the D1S interaction, the same
shown in Fig.~\ref{fig:dox2-}, are now compared with those obtained by
using the D1ST force, indicated here by the dashed lines.  In this
figure we show the full spectrum by including the DRPA results below
the continuum threshold, and the CRPA results above it. We observe
that the strengths are concentrated in three different regions.  A
first one between 10 and 15 MeV, a second one between 16 and 20 MeV,
and a third one around 25 MeV.

The excitations between 10 and 15 MeV are dominated by the
$[(1d_{5/2})(1p_{1/2})^{-1}]$ proton transitions. The calculations
with and without tensor predict almost identical excitation energies
for these resonances in the $^{16}$O and $^{28}$O nuclei, while for
the $^{22}$O and $^{24}$O nuclei the resonances of the calculations
with D1ST have larger energies than those obtained with D1S. This is a
consequence of the Otsuka effect of the tensor force on the
s.p. energies. In the $^{22}$O and $^{24}$O, because of the occupancy
of the neutron $(1d_{5/2})^{-1}$ s.p. level, the energy of the proton
$(1p_{1/2})^{-1}$ level is lowered while that of the $(1d_{5/2})$
one is enhanced. The transition between these two states requires
more energy and this implies an increase of the excitation energy of
the nucleus. Because all the s.p. spin-orbit partner levels are
occupied in $^{16}$O and $^{28}$O nuclei, the effect we have just
described is always compensated by a similar one of different
sign. For this reason, in these latter nuclei, the energies obtained
in calculations with and without tensor almost coincide.  The
resonances observed at higher energies are composed by various p-h
excitations and it is difficult to identify a dominant transition.
The strength below 10 MeV in the three heavier oxygen isotopes is
produced by excited states dominated by the
$[(2p_{3/2})(1d_{5/2})^{-1}]$ neutron transition.

Experimentally, the $2^-$ excitation of $^{16}$O has been studied with
electron scattering \cite{kue83}. In this investigation a total \bmtw
strength of $1052 \pm 272$ \mun fm$^2$ has been found. Our
calculations predict larger values, around $1860$ \mun fm$^2$. In
Refs. \cite{ram91,neu99} results of RPA calculations predicting total
strengths of $1204$ \mun fm$^2$ are quoted. We have calculated
the $2^-$ excitation by using a phenomenological model with the
Landau-Migdal residual interaction proposed in Ref.  \cite{co90}, and
we have obtained a total strength of $1670$ \mun fm$^2$. 

%%%%%%%%%%%%%%%%%%%%%%%%%%%%%%%%%%%%%%%%%%%%%%%%%%%%%%%%%%%%%%%%%%%%%%
% oxy Discrete 3+
%%%%%%%%%%%%%%%%%%%%%%%%%%%%%%%%%%%%%%%%%%%%%%%%%%%%%%%%%%%%%%%%%%%%%%
\begin{figure}[ht]
\begin{center}
\includegraphics [scale=0.6]{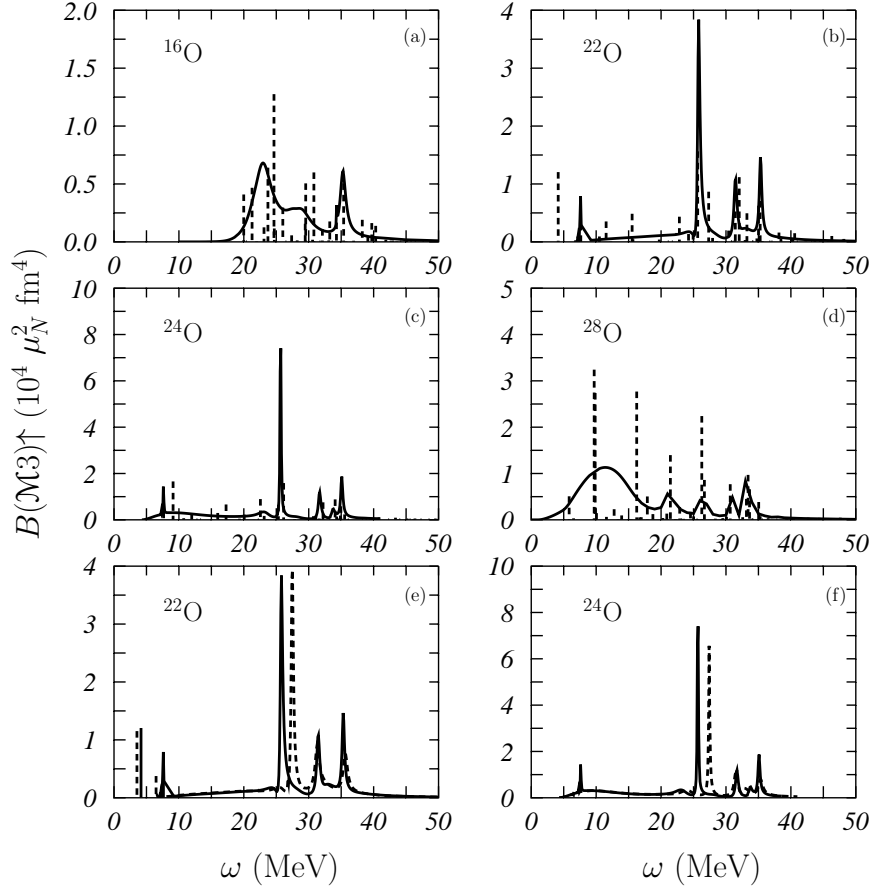}
\caption{\small \bmth values for the oxygen isotopes under
  investigation as a function of the excitation energy. In panels (a)-(d)
  we compare DRPA, vertical lines, and CRPA, full lines, 
  results obtained with the D1S interaction.
  For the $^{22}$O  and $^{24}$O isotopes only, we compare,   
  in the panels (e) and (f), the CRPA results obtained 
  with the D1S and D1ST interactions indicated, respectively, by the  
  full and dashed curves. In these two latter panels we added to the
  CRPA results also the DRPA results obtained below the emission
  particle threshold.  
}
\label{fig:dox3+}
\end{center}
\end{figure}

The investigation done for the $2^-$ excitation has been repeated for
the $3^+$ mode. We summarize in Fig.~\ref{fig:dox3+} our results
regarding the energy distribution of the \bmth strength for the oxygen
isotopes. In the panels (a), (b), (c) and (d) we compare DRPA
(dashed vertical lines) and CRPA (full lines) calculations
done with the D1S interaction. In the other two panels we compare the
RPA results obtained with the D1S (full lines) and the D1ST (dashed
curves) interactions for the $^{22}$O and $^{24}$O isotopes. For the
other two oxygen isotopes under investigation the results obtained
with and without tensor almost overlap, as observed for the $1^+$ and
$2^-$ cases.  In these two lower panels, we present the full
excitation spectrum obtained by adding to the CRPA results also those
obtained with the DRPA approach in the region below the continuum
threshold.

The results of Fig.~\ref{fig:dox3+} indicate a good agreement between
the position of the peaks obtained in discrete and continuum
calculations. With the exception of the $^{28}$O nucleus, the main
part of the strength is located above 20 MeV. The main peaks are
dominated by the $[(1f_{7/2})(1p_{1/2})^{-1}]$ proton transition. The
difference between the positions of the peaks obtained with and
without tensor is due mainly to the Otsuka effect. In these two nuclei
the occupation of the neutron $(1d_{5/2})^{-1}$ level decreases the
energy value of the proton $(1p_{1/2})^{-1}$ level, and enhances that
of the $(1f_{7/2})$ one. The increase of the bare p-h energy
difference induces a difference in the position of the peaks obtained
in the CRPA calculations.

%%%%%%%%%%%%%%%%%%%%%%%%%%%%%%%%%%%%%%%%%%%%%%%%%%%%%%%%%%%%%%%%%%%%%%
% Ca 2-
%%%%%%%%%%%%%%%%%%%%%%%%%%%%%%%%%%%%%%%%%%%%%%%%%%%%%%%%%%%%%%%%%%%%%%
\begin{figure}[b]
\begin{center}
\includegraphics [scale=0.6]{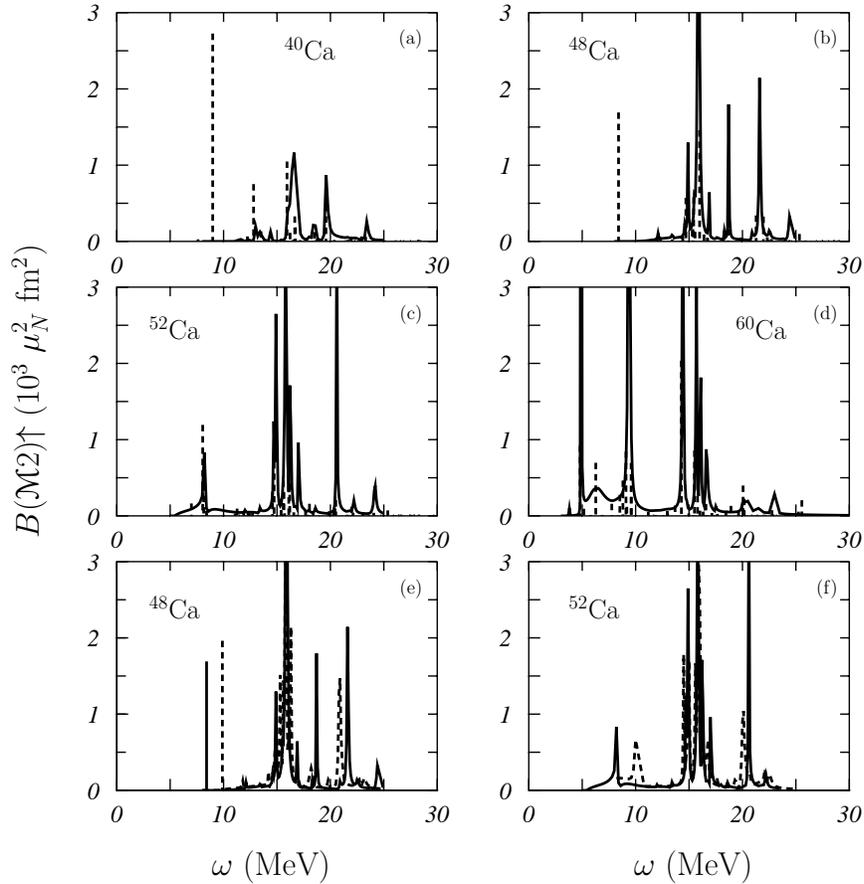}
\caption{\small The same as Fig.~\ref{fig:dox3+}. In this case we
  present the results of the \bmtw values for the calcium isotopes under
  investigation. 
}
\label{fig:dca2-}
\end{center}
\end{figure}

The study on the $2^-$ and $3^+$ has been carried on also for the
calcium isotopes and we show in Fig. \ref{fig:dca2-} the energy
distribution of the \bmtw excitation in these nuclei.  The structure
of the figure is analogous to that of Fig.~\ref{fig:dox3+}. In the
four upper panels we compare discrete and continuum RPA results
obtained with the D1S interaction and in the two lower panels we show
the effect of the tensor force for the two isotopes where this effect
is remarkable.  As it has been done for the oxygen cases, in these
last two panels we show the full spectrum obtained by adding the DRPA
results obtained below the continuum threshold to the CRPA results
obtained above it.

As always observed in the previous cases, we remark a good agreement
between the position of the peaks obtained in discrete and continuum
results.  In all the responses we identify a sharp excitation at
energies below 10 MeV and wider, and more fragmented, excitations
around 15 MeV.  

The excitation below 10 MeV is dominated by a specific p-h excitation,
the $[(1f_{7/2})(1d_{3/2})^{-1}]$ proton transition. The energy of
this excitation is below the particle emission threshold for the
$^{40}$Ca and $^{48}$Ca nuclei. The comparison of the results obtained
with and without tensor can be explained by means of the Otsuka effect
absent in the $^{40}$Ca and $^{60}$Ca isotopes.  In $^{48}$Ca and
$^{52}$Ca, where the neutron $(1f_{7/2})^{-1}$ state is occupied while
its spin-orbit partner level is empty, the tensor force, acting
between this state and the proton states, lowers the energy of the
proton $(1d_{3/2})^{-1}$ level, and enhances that of the $(1f_{7/2})$
one, increasing the energy difference between these two levels, with
the obvious consequences on the nuclear excitation energies evident in
the panels (e) and (f) of the figure.

The resonances at about 15 MeV have more collective character.  In all
the isotopes we have considered, these resonances are dominated by the
same proton and neutron s.p. excitations, which are not related to the
neutron excess of some of the isotopes considered.  This indicates a
collective common feature of all the nuclei under consideration.

An experimental study of the $2^-$ excitation in the $^{48}$Ca nucleus
has been carried on with electron scattering \cite{neu99}.  In this
work, the study of the \bmtw energy distribution has been limited to
excitation energies smaller than 15 MeV (see Fig. 2 of
Ref.~\cite{neu99}).  The experimental strength shows a sharp peak at
about 8 MeV, and broader structures at 12 and 15 MeV.  The
experimental value of the energy weighted sum rule deduced from Fig. 3
of Ref. \cite{neu99} is about $(17 \pm 2) \cdot 10^3$ \mun MeV fm$^2$.
By integrating up 16 MeV our \bmtw continuum strengths we obtain, for
the energy weigthed sum rule, the values of 29.89 and 24.87 in $10^3$
\mun MeV fm$^2$ units, for the D1S and D1ST interactions, respectively.
Extending the integral up to 30 MeV we obtain, respectively, 59.71 and
53.54 in $10^3$ \mun MeV fm$^2$ units. These values are in agreement
with the RPA results of $52.4 \cdot 10^3$ \mun MeV fm$^2$, presented
in Ref.~\cite{neu99}.  In this reference it is shown that an
improvement of the description of the experimental data is obtained
when 2p-2h excitations are considered in RPA calculations.  In any
case, we should remark that the position of the peaks is not changed
by the 2p-2h calculations. 

%%%%%%%%%%%%%%%%%%%%%%%%%%%%%%%%%%%%%%%%%%%%%%%%%%%%%%%%%%%%%%%%%%%%%%
% Ca 3+ 
%%%%%%%%%%%%%%%%%%%%%%%%%%%%%%%%%%%%%%%%%%%%%%%%%%%%%%%%%%%%%%%%%%%%%%
\begin{figure}[ht]
\begin{center}
\includegraphics [scale=0.6]{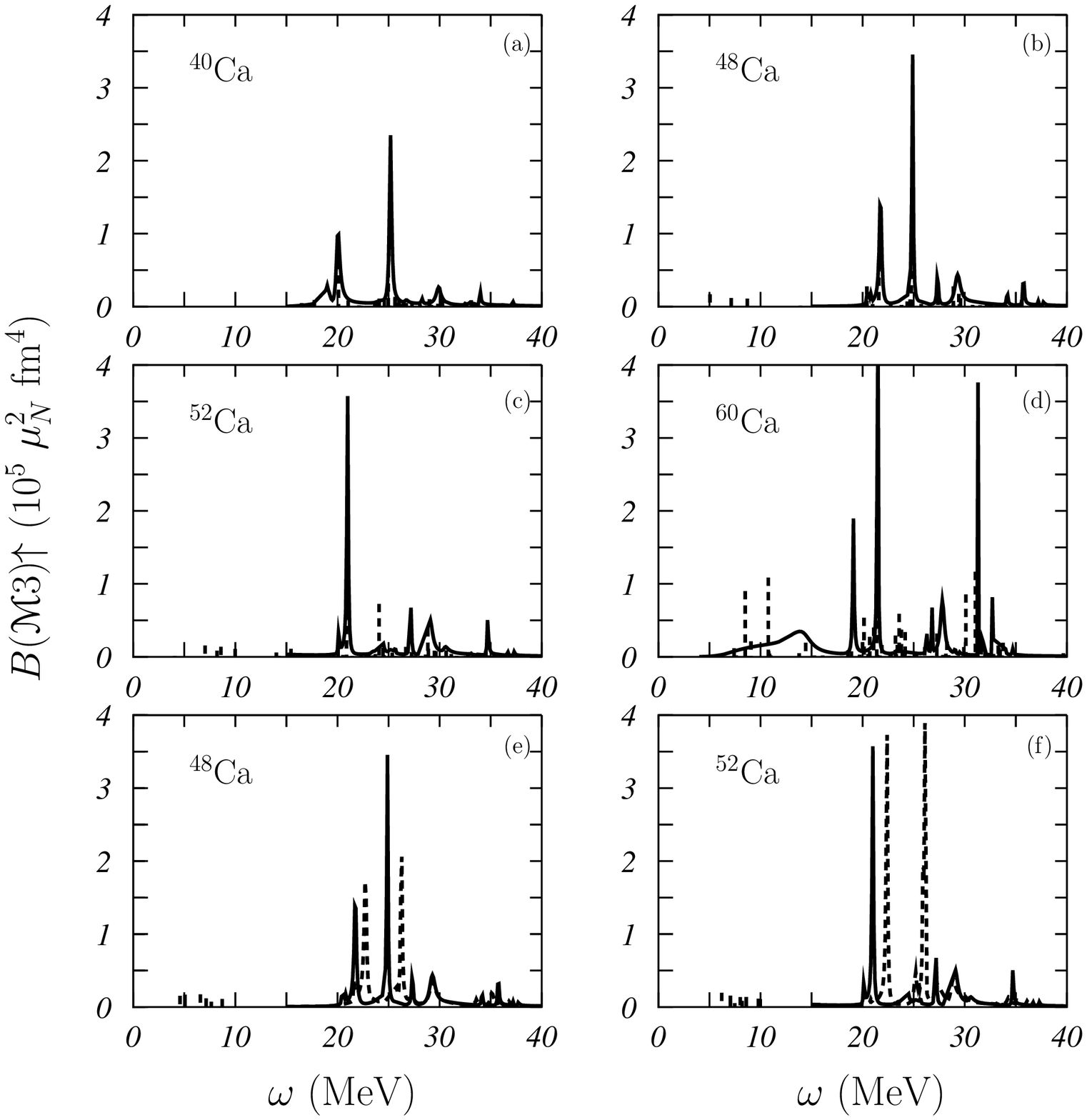}
\caption{\small The same as Fig. \ref{fig:dca2-} for the \bmth values. 
}
\label{fig:dca3+}
\end{center}
\end{figure}

The results of our calculations for the $3^+$ excitations in calcium
isotopes are shown in Fig. \ref{fig:dca3+} whose structure is
analogous to that of the previous two figures.  Also in this case, we
remark the agreement between the position of the peaks obtained in the
discrete and continuum RPA calculations. The size of the \bmth
strengths is analogous in all the isotopes considered.  The strengths
are mainly concentrated above 20 MeV, indicating that they are mainly
generated by $2 \hbar \omega$ excitations, i.e. transition between
levels belonging to two major shells.  In the three heavier isotopes,
some strength below 10 MeV is present. This is produced by the $0
\hbar \omega$ excitations of neutrons occupying the $f$ s.p. levels. 

Also in this case, the tensor force has no effects on the responses of
the two nuclei where all the spin-orbit partner levels are occupied,
i. e. $^{40}$Ca and $^{60}$Ca.  For this reason, in the two lower
panels, we show only the results related to $^{48}$Ca and
$^{52}$Ca, where we found a remarkable effect. A detailed
study of the resonances indicates that the shift between the peak
positions is mainly due to the Otsuka effect.

\section{Conclusions}   
\label{sec:conclusions}
In this article we have presented results of discrete and continuum
self-consistent RPA calculations of magnetic excitations of some
spherical oxygen and calcium isotopes. The calculations have been done
by using the finite-range D1S and D1M interactions of Gogny type
\cite{ber89,ber91,gor09} and other two new parametrizations, D1ST and
D1MT \cite{ang11}, obtained by adding a tensor-isospin to them.

As expected, the more striking result emerging from our calculations is
related to the $1^+$ excitation.  The strengths of this excitation in
nuclei where all the spin-orbit partner levels are occupied are orders
of magnitude smaller than those in nuclei where one of the spin-orbit
partner levels is empty. In this latter group of nuclei, a spin-flip
transition between the two spin-orbit partner levels is allowed. 
This difference between the responses of the two groups of
nuclei is typical of the $1^+$ excitation. The sizes of the $2^-$ and
$3^+$ responses have similar magnitudes in all the nuclei
investigated, even in the $3^+$ states where the same spin-flip
transitions dominating the $1^+$ excitation are present.

Our study indicates that, in general, magnetic excitations are more
related to the s.p. structure of the nucleus than to collective
effects. We found some indications of collectivity in the $2^-$ and
$3^+$ responses around 20-25 MeV in the oxygen isotopes, and at
slightly smaller energies in the calcium isotopes. In any case, the
main magnetic excitations are dominated by s.p. transitions. This does
not mean that the effect of the residual interaction is negligible. The
residual interaction moves the peak positions with respect to those of
the IPM, and adds to the main p-h component also the contributions of
other p-h transitions. This weaker components change the RPA wave
function, and produce effects which show up at high momentum transfer
values, above 2 fm$^{-1}$. 

Since in magnetic excitations the residual interaction produces
effects which can be treated as perturbations with respect to the main
IPM response, it is possible to isolate the role of the various terms
of the interaction. This feature has allowed a detailed investigation
of the tensor-isospin term of the residual interaction. We have
observed that relevant tensor effects are present only in those nuclei
where not all the spin-orbit partner levels are occupied.  In these
nuclei, the effects of the tensor term of the interaction are active
in both HF and RPA calculations. The tensor force changes the values
of the s.p. energies. The effect pointed out by Otsuka and
collaborators \cite{ots05,ots06} affects nucleons of different type.
In our calculations, we found a similar effect acting also between
nucleons of the same type.  We have identified the consequences of the
genuine Otsuka effect, and of its analogous, in $^{22}$O, $^{24}$O,
$^{48}$Ca, $^{52}$Ca nuclei. In addition to these effects on the
s.p. energies and wave functions, the tensor force affects also the
RPA calculations, mainly in those nuclei where not all the spin-orbit
partner levels are occupied. 

In our calculations, the effects of the tensor force act against those
of the other terms of the interaction. We have discussed in detail the
case of the $1^+$ excitation, where the residual interaction moves the
position of the main peaks at higher energies with respect to the IPM
results. The modification of the s.p. energies due to the tensor
lowers the value of the excitation energy, and the presence of the
tensor in the RPA calculation further diminishes this value which at
the end results to be close to the original IPM value.  Furthermore,
the study of the RPA wave functions in the peak position, used to
calculate the electron scattering transverse responses, indicates that
the presence of the tensor decreases the contribution of the p-h
components different from the dominant one.

The comparison with the few experimental data available indicates that
our calculations describe reasonably well the position of the excited
states but they overestimate the strength of the magnetic
excitations. For example, the values of the \bmon strengths in
$^{48}$Ca are more than two times larger than those indicated by the
experiment. This result is common to all the mean-field shell model
and RPA calculations, and it is known in the literature as the
quenching problem \cite{ram91}. Also the comparison with the $2^-$
excitation data in $^{48}$Ca \cite{neu99} indicates the need of
quenching. We should remark, however, that, in this case, we found a
non negligible amount of strength beyond the maximum excitation energy
explored by the experiment.  We have pointed out, by comparing our
$1^+$ electron scattering responses in $^{48}$Ca with the experimental
one \cite{ste83}, the need of a momentum dependent quenching factor to
reproduce the data.  This indicates that the physics behind the
quenching effects is rather involved, and cannot be simulated by a
simple reduction factor.

Another indication of the need of extending the traditional RPA
approach is coming from the experimental observation of non negligible
\bmon strength in $^{16}$O and $^{40}$Ca \cite{kue83,gro79}.  We have
already mentioned that, in our approach, the \bmon strength is
negligible for nuclei with all spin-orbit partner levels occupied.
Experimentally, a \bmon strength of about 1.0 \mun has been identified
in $^{16}$O \cite{kue83} and of about 1.2 \mun in $^{40}$Ca
\cite{gro79}. The description of these non-negligible strengths have
been explained by considering 2p-2h excitations \cite{ari80,kam04}.

The study of magnetic excitations in medium-heavy nuclei having a
relatively simple structure offers the opportunity to obtain
information about some basic nuclear effects, such as shell closure,
tensor force, spin and orbital terms of the electromagnetic operator,
and correlations.  The new radioactive ion beams accelerators allow
the production of unstable nuclei such as those we have investigated
in our paper. The further implementation of electron scattering
facility \cite{sud09,sud11,ant11} would be extremely useful for this
type of investigations. 

\acknowledgments 
This work has been partially supported by the PRIN (Italy) {\sl
Struttura e dinamica dei nuclei fuori dalla valle di stabilit\`a}, by
the Spanish Ministerio de Ciencia e Innovaci\'on (Contract
Nos. FPA2009-14091-C02-02 and ACI2009-1007) and by the Junta de
Andaluc\'{\i}a (Grant No. FQM0220).

%

%%%%%%%%%%%%%%%%%%%%%%%%%%%%%%%%%%%%%%%%%%%%%%%%%%%%%%%%%%%%%%%%%%%%%%
%%%%%%%%%%%%%%%%%%%%%%%%%%%%%%%%%%%%%%%%%%%%%%%%%%%%%%%%%%%%%%%%%%%%%%
%%%%%%%%%%%%%%%%%%%%%%%%%%%%%%%%%%%%%%%%%%%%%%%%%%%%%%%%%%%%%%%%%%%%%%
%%%%%%%%%%%%%%%%%%%%%%%%%%%%%%%%%%%%%%%%%%%%%%%%%%%%%%%%%%%%%%%%%%%%%%
%%%%%%%%%%%%%%%%%%%%%%%%%%%%%%%%%%%%%%%%%%%%%%%%%%%%%%%%%%%%%%%%%%%%%%


\begin{thebibliography}{10}
\expandafter\ifx\csname url\endcsname\relax
  \def\url#1{\texttt{#1}}\fi
\expandafter\ifx\csname urlprefix\endcsname\relax\def\urlprefix{URL }\fi

\bibitem{don11a}
V.~De~Donno, G.~Co', M.~Anguiano, A.~M. Lallena, Phys. \ Rev. \ C 83 (2011)
  044324.

\bibitem{don11b}
V.~De~Donno, M.~Anguiano, G.~Co', A.~M. Lallena, Phys. \ Rev. \ C 84 (2011)
  037306.

\bibitem{bro06}
B.~A. Brown, T.~Duguet, T.~Otsuka, D.~Abe, T.~Suzuki, Phys. \ Rev. \ C 74
  (2006) 061303(R).

\bibitem{les07}
T.~Lesinski, M.~Bender, K.~Bennaceur, T.~Duguet, J.~Meyer, Phys. \ Rev. \ C 76
  (2007) 014312.

\bibitem{col07}
G.~Col\`o, H.~Sagawa, S.~Fracasso, P.~F. Bortignon, Phys. \ Lett. \ B 646
  (2007) 227.

\bibitem{bri07}
D.~M. Brink, F.~Stancu, Phys. \ Rev. \ C 75 (2007) 064311.

\bibitem{col08}
G.~Col\`o, H.~Sagawa, S.~Fracasso, P.~F. Bortignon, Phys. \ Lett. \ B 668
  (2008) 457.

\bibitem{ben09}
M.~Bender, K.~Bennaceur, T.~Duguet, P.~H. Heenen, T.~Lesinski, J.~Meyer, Phys.
  \ Rev. \ C 80 (2009) 064302.

\bibitem{mor10}
M.~Moreno-Torres, M.~Grasso, H.~Liang, V.~De~Donno, M.~Anguiano, N.~van Giai,
  Phys. \ Rev. \ C 81 (2010) 064327.

\bibitem{bai09a}
C.~L. Bai, H.~Sagawa, H.~Q. Zhang, X.~Z. Zhang, G.~Col\`o, F.~R. Xu, Phys. \
  Lett. \ B 675 (2009) 28.

\bibitem{bai09b}
C.~L. Bai, H.~Q. Zhang, X.~Z. Zhang, F.~R. Xu, H.~Sagawa, G.~Col\`o, Phys. \
  Rev. \ C 79 (2009) 041301(R).

\bibitem{cao09}
L.-G. Cao, G.~Col\`o, H.~Sagawa, P.~F. Bortignon, L.~Sciacchitano, Phys. \ Rev.
  \ C 80 (2009) 064304.

\bibitem{bai10}
C.~L. Bai, H.~Q. Zhang, H.~Sagawa, X.~Z. Zhang, G.~Col\`o, F.~R. Xu, Phys. \
  Rev. \ Lett. 105 (2010) 072501.

\bibitem{dec80}
J.~Decharg\`e, D.~Gogny, Phys. \ Rev. \ C 21 (1980) 1568.

\bibitem{ber89}
J.~F. Berger, M.~Girod, D.~Gogny, Nucl. \ Phys. \ A 502 (1989) 85c.

\bibitem{ber91}
J.~F. Berger, M.~Girod, D.~Gogny, Comp. \ Phys. \ Commun. 63 (1991) 365.

\bibitem{cha07t}
F.~Chappert, Nouvelles param\'etrisation de l'interaction nucl\'eaire effective
  de Gogny, Ph.D. thesis, Universit\'e de Paris-Sud XI (France),
  http://tel.archives-ouvertes.fr/tel-001777379/en/ (2007).

\bibitem{cha08}
F.~Chappert, M.~Girod, S.~Hilaire, Phys. \ Lett. \ B 668 (2008) 420.

\bibitem{gor09}
S.~Goriely, S.~Hilaire, M.~Girod, S.~P\'eru, Phys. \ Rev. \ Lett. 102 (2009)
  242501.

\bibitem{ang11}
M.~Anguiano, G.~Co', V.~De~Donno, A.~M. Lallena, Phys. \ Rev. \ C 83 (2011)
  064306.

\bibitem{sil06}
T.~Sil, S.~Shlomo, B.~K. Agrawal, P.~G. Reinhard, Phys. \ Rev. \ C 73 (2006)
  034316.

\bibitem{per05}
S.~P\'eru, J.~F. Berger, P.~F. Bortignon, Eur. \ Phys. \ J. \ A 26 (2005) 25.

\bibitem{edm57}
A.~R. Edmonds, Angular momentum in quantum mechanics, Princeton University
  Press, Princeton, 1957.

\bibitem{rin80}
P.~Ring, P.~Schuck, The nuclear many-body problem, Springer, Berlin, 1980.

\bibitem{ama93}
J.~E. Amaro, G.~Co', A.~M. Lallena, Ann. \ Phys. \ (N.Y.) 221 (1993) 306.

\bibitem{deh82t}
R.~de~Haro, Analysis of nuclear excited states. the escape and sprading widths,
  Ph.D. thesis, Rheinische Friedrich-Wilhelms-Universit{\"a}t Bonn, Germany,
  unpublished (1982).

\bibitem{deh82}
R.~de~Haro, S.~Krewald, J.~Speth, Nucl. \ Phys. \ A 388 (1982) 265.

\bibitem{co06b}
G.~Co', Acta. \ Phys. \ Pol. \ B 37 (2006) 2235.

\bibitem{suh07}
J.~Suhonen, From nucleons to nucleus, Springer, Berlin, 2007.

\bibitem{don08t}
V.~De~Donno, Nuclear excited states within the random phase approximation
  theory, Ph.D. thesis, Universit\`a del Salento (Italy),
  http://www.fisica.unisalento.it/~gpco/stud.html (2008).

\bibitem{co09b}
G.~Co', V.~De~Donno, C.~Maieron, M.~Anguiano, A.~M. Lallena, Phys. \ Rev. \ C
  80 (2009) 014308.

\bibitem{don09}
V.~De~Donno, G.~Co', C.~Maieron, M.~Anguiano, A.~M. Lallena, M.~Moreno-Torres,
  Phys. \ Rev. \ C 79 (2009) 044311.

\bibitem{akm98}
A.~Akmal, V.~R. Pandharipande, D.~G. Ravenhall, Phys. \ Rev. \ C 58 (1998)
  1804.

\bibitem{gan10}
S.~Gandolfi, A.~Y. Illarionov, S.~Fantoni, J.~C. Miller, F.~Pederiva, K.~E.
  Schmidt, Mont. \ Not. \ R. \ Astron. \ Soc. 404 (2010) L35.

\bibitem{del10}
J.-P. Delaroche, M.~Girod, J.~Libert, H.~Goutte, S.~Hilaire, S.~P\'eru,
  N.~Pillet, G.~F. Bertsch, Phys. \ Rev. \ C 81 (2010) 014303.

\bibitem{co11b}
G.~Co', V.~De~Donno, P.~Finelli, M.~Grasso, M.~Anguiano, A.~M. Lallena,
  C.~Giusti, A.~Meucci, F.~D. Pacati, Phys. \ Rev. \ C 85 (2012) 024322.

\bibitem{hei83}
J.~Heisenberg, H.~Blok, Ann. \ Rev. \ Nucl. \ Par. \ Sci. 33 (1983) 569.

\bibitem{ots05}
T.~Otsuka, T.~Suzuki, R.~Fujimoto, H.~Grawe, Y.~Akaishi, Phys. \ Rev. \ Lett.
  95 (2005) 232502.

\bibitem{ots06}
T.~Otsuka, T.~Matsuo, D.~Abe, Phys. \ Rev. \ Lett. 97 (2006) 162501.

\bibitem{ste80}
W.~Steffen, H.-D. Gr{\"a}f, W.~Gross, D.~Meuer, A.~Richter, E.~Spamer,
  O.~Titze, W.~Kn{\"u}pfer, Phys. \ Lett. \ B 95 (1980) 23.

\bibitem{ste83}
W.~Steffen, H.-D. Gr{\"a}f, A.~Richter, A.~H{\"a}rting, W.~Weise,
  U.~Deutchmann, G.~Lahm, R.~Neuhausen, Nucl. \ Phys. \ A 404 (1983) 413.

\bibitem{knu83}
W.~Kn{\"u}pfer, B.~C. Metsch, A.~Richter, Phys. \ Lett. \ B 129 (1983) 375.

\bibitem{kue83}
G.~K{\"u}chler, A.~Richter, E.~Spamer, W.~Steffen, W.~Kn{\"u}pfer, Nucl. \
  Phys. A 406 (1983) 473.

\bibitem{ram91}
S.~Raman, L.~W. Fagg, R.~S. Hicks, Electric and magnetic giant resonances in
  nuclei, {\rm J. Speth ed.}, World Scientific, Singapore, 1991.

\bibitem{neu99}
P.~von Neumann-Cosel, F.~Neumeyer, S.~Nishizaki, V.~yu. Ponomarev,
  C.~Rangacharyulu, B.~Reitz, A.~Richter, G.~Schrieder, D.~I. Sober,
  T.~Weindzoch, J.~Wambach, Phys. \ Rev. \ Lett. 82 (1999) 1105.

\bibitem{co90}
G.~Co', A.~M. Lallena, Nucl. \ Phys. \ A 510 (1990) 139.

\bibitem{gro79}
W.~Gross, D.~Meuer, A.~Richter, E.~Spamer, O.~Titze, W.~Kn{\"u}pfer, Phys. \
  Lett. \ B 84 (1978) 296.

\bibitem{ari80}
A.~Arima, D.~Strottman, Phys. \ Lett. \ B 96 (1980) 23.

\bibitem{kam04}
S.~Kamerdzhiev, J.~Speth, G.~Tertychny, Phys. \ Rep. 393 (2004) 1.

\bibitem{sud09}
T.~Suda, et~al., Phys. \ Rev. \ Lett. 102 (2009) 102501.

\bibitem{sud11}
T.~Suda, Journal \ of \ Physics: \ Conference \ Series 267 (2011) 012088.

\bibitem{ant11}
A.~N. Antonov, et~al., Nucl. \ Inst. \ Met. \ A 637 (2011) 60.

\end{thebibliography}
\end{document}